\documentclass[reprint, aps, prapplied,preprintnumbers, floatfix, longbibliography]{revtex4-2}
\usepackage[utf8]{inputenc}
\usepackage{amsmath}
\usepackage{amssymb}
\usepackage{verbatim}
\usepackage{subcaption}
\usepackage{graphicx}
\usepackage{float}
\usepackage{booktabs}
\usepackage{wrapfig}
\usepackage{xcolor}
\usepackage{mathtools}
\usepackage{hyperref}

\definecolor{colorblue}{RGB}{4,4,236}
\hypersetup{
    colorlinks=true,  
    linkcolor=colorblue,   
    citecolor=colorblue,    
    urlcolor=colorblue, 
    filecolor=colorblue   
}

\begin{document}  
\preprint{APS/123-QED}\preprint{FERMILAB-PUB-24-0824-T}

\title{Uniform Field in Microwave Cavities Through the Use of Effective Magnetic Walls}

\author{Jim~A.~Enriquez$^{\dagger}$}
\thanks{Email: jaenriqueze@unal.edu.co}
\affiliation{Physics Department, National University of Colombia, 111321, Bogota, Colombia}

\author{Rustam~Balafendiev}
\thanks{These authors contributed equally.}
\affiliation{Science Institute, University of Iceland, Dunhagi 5, 107 Reykjavik, Iceland}

\author{Alexander~J.~Millar}
\affiliation{Theoretical Physics Division, Fermi National Accelerator Laboratory, Batavia, IL 60510, USA}
\affiliation{Superconducting Quantum Materials and Systems Center (SQMS), Fermi National Accelerator Laboratory, Batavia, IL 60510, USA}

\author{Constantin~Simovski}
\affiliation{Department of Electronics and Nanoengineering, Aalto University, Maarintie 8, FI00076, Espoo, Finland}
\author{Pavel~Belov}
\affiliation{Qingdao Innovation and Development Center, Harbin Engineering University, Qingdao 266000, Shandong, China}
\affiliation{School of Engineering, New Uzbekistan University, Movarounnahr str. 1, 100000, Tashkent, Uzbekistan}

\date{\today}

\graphicspath{{./figures}}

\makeatletter
\newcommand{\Jim}[1]{\textcolor{violet}{#1}}
\newcommand{\Rustam}{\@ifstar{\@Ra}{\@Rb}}
\newcommand{\@Rb}[1]{\textcolor[HTML]{0da34e}{#1}}
\newcommand{\@Ra}[1]{\textcolor[HTML]{0da34e}{\textbf{Rustam:} #1}}
\begin{abstract}
Wire media (WM) resonators have emerged as promising realization for plasma haloscopes -- devices designed to detect axions, a potential component of dark matter. Key factors influencing the detection probability include cavity volume, resonance quality factor, and form factor. While the form factor has been explored for resonant frequency tuning, its optimization for axion detection remains unexplored. In this work, we present a novel approach to significantly enhance the form factor of WM plasma haloscopes. By shifting the metal walls of the resonator by a quarter wavelength, we effectively convert an electric wall boundary condition into a magnetic wall one, allowing for an almost uniform mode. Theoretical analysis and numerical simulations confirm that this modification improves the electric field profile and boosts the form factor, while also slightly enhancing the quality factor. We validate these findings through experimental results from two prototype resonators: one with a standard geometry and another with a quarter-wave air gap between the WM and the walls. Additionally, our method provides a simple way to control the field profile within WM cavities, which can be explored for further applications.
\end{abstract}

\maketitle

\section{Introduction}
 \begin{figure*}[t]
\centering
\includegraphics[width=1\linewidth]{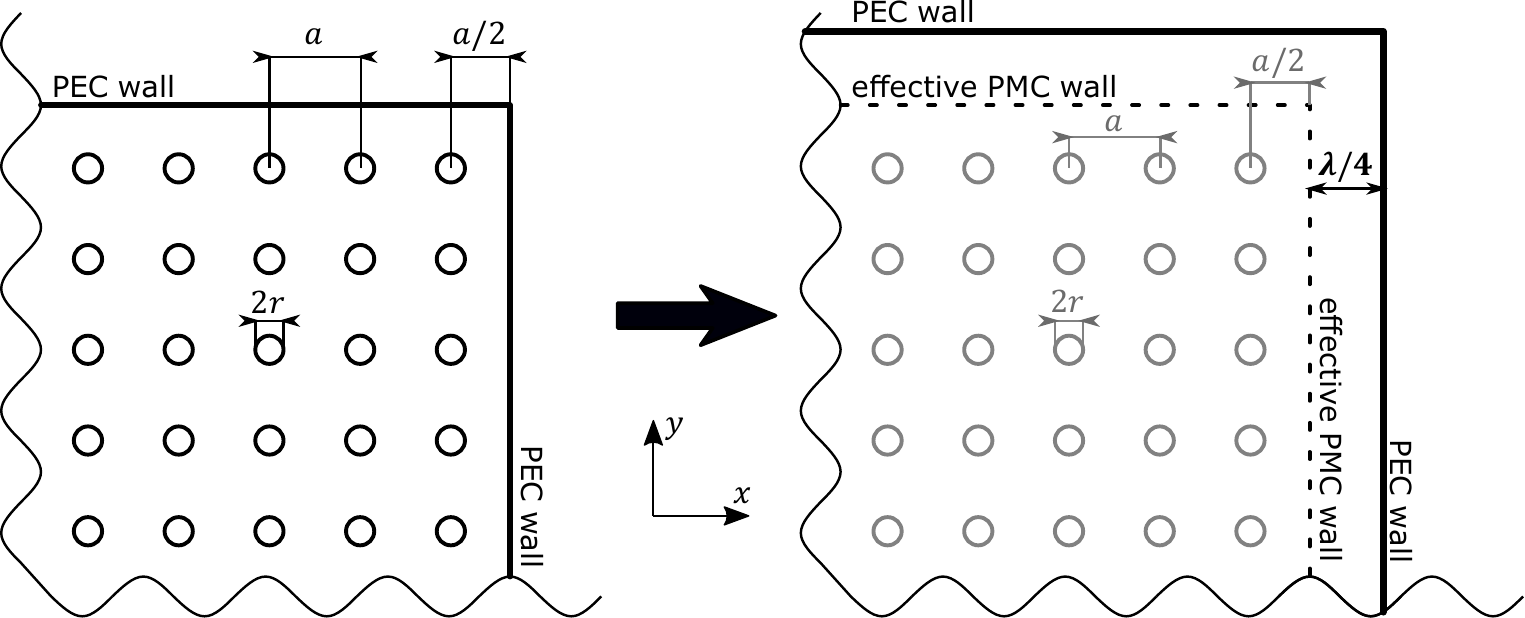}
    \caption{Schematic representation of the proposed modification to the cavity's boundary condition. A wire medium (WM) resonator consists on an arrange of metallic wires with radius $r$, distributed with a period $a$. In a regular configuration, the distance between the last set of wires and the cavity walls, made of perfect electric conductor (PEC), is $a/2$. By adjusting this distance to $a/2+\lambda/4$, the boundary enclosing the WM behaves effectively as a perfect magnetic conductor (PMC) wall.}
    \label{fig:sch}
\end{figure*}
Wire media (WM), initially proposed by Brown \cite{Brown} and Rotman \cite{Rotman}, and later studied by Pendry and Belov et al. in works \cite{pendry_low_1998,Belov}, have garnered significant attention for their distinctive properties as an artificial plasma material. These structures, comprising a dense array of parallel metallic wires embedded in a dielectric host, exhibit a tunable plasma frequency that can be adjusted by modifying their geometrical parameters. Beyond theoretical exploration, WM-based devices have found practical applications in diverse fields, ranging from  microwave antennas \cite{forati_epsilon-near-zero_2015}  to near-field thermophotovoltaic systems \cite{OE} and systems of radiative cooling \cite{simovski_hyperlens_2015}. A detailed review about WM applications can be found in Ref.~\cite{simovski2012wire}. 

Previous investigations into WM open resonators have primarily focused on exploiting topological transitions in their isofrequency contours to achieve a notable Purcell factor \cite{mirmoosa_magnetic_2016, mirmoosa_double_2016}. More recently, cavities consisting in WM enclosed by metallic walls and placed in a DC magnetic field, often referred to as plasma haloscopes \cite{lawson_tunable_2019}, have emerged as a promising technique for the detection of axions, a possible candidate for dark matter \cite{ALPHA:2022rxj}. One can also look for dark photons, another promising dark matter candidate, even if the magnetic field is absent~\cite{Gelmini:2020kcu}. Additionally, the detection of high-frequency gravitational waves can be explored \cite{Gatti2024, capdevilla2024gravitational}. Our earlier studies of WM-filled cavities demonstrated that WM operating in the epsilon-near-zero (ENZ) regime significantly extends the wavelength of light within the medium, resulting in a low-frequency fundamental resonance and a high degree of field uniformity \cite{Balafendiev2022}. 

The improvement of field homogeneity in microwave resonators have been studied in the context of electron paramagnetic resonance (EPR), where axially uniform field distribution for a TE mode is a desirable feature \cite{hyde_uniform_2019}. A common approach to higher uniformity of the field pattern in hollow metal cavities involves a design with a central region surrounded by two end regions. The central region functions as a metallic waveguide operating at its cutoff frequency, while the optical length of the end regions is set to a quarter of a wavelength \cite{hyde_uniform_2019}. 
Various configurations for end regions in hollow metal cavities have been proposed, including dielectric rods with the same cross-section as the central region \cite{mett_axially_2001}, modified cross sections larger than that of the central region \cite{anderson_cavities_2002}, and conducting rods insertions \cite{hyde_cavities_2002}.

When the axion frequency matches the resonant frequency of a microwave cavity mode, the power generated by axion-photon conversion is given by \cite{Sikivie1983,ALPHA:2022rxj}:
 
\begin{equation}
    P_a\approx g_{a\gamma \gamma}^2\frac{\rho_a}{m_a}B_0^2VC_{lmn}\mathrm{min}(Q_L,Q_a)
\end{equation}
where $g_{a\gamma \gamma}$ is the axion-photon coupling constant, $m_a$ is the axion mass, $\rho_a$ is the local mass density of the axion field, and $\mathbf{B}_0$ is the external DC magnetic field.  The parameters $V$ and $Q_L$ denote the cavity volume and loaded quality factor, respectively, while $Q_a$ represents the quality factor of the galactic halo axion signal. $C_{lmn}$ is the form factor associated with the $\mathrm{TM}_{lmn}$ mode of the cavity. To maximize the axion-photon conversion power in axion haloscopes, it is crucial to optimize the external DC magnetic field, cavity volume, quality factor, and form factor. Notably, the scanning rate of an axion search experiment, which probes a range of axion masses, is proportional to the square of the form factor \cite{asztalos_large-scale_2001}.   

For axion haloscopes, field uniformity is as crucial as it is for EPR devices, since the form factor directly quantifies this uniformity. The form factor for a mode in a plasma haloscope, characterized by an electric field $\mathbf{E}_{lmn}$, is defined as
 \begin{equation}\label{Ffactor}
         C_{lmn} = \frac{\left|\iiint_V \mathbf{B}_0 \cdot \mathbf{E}_{lmn} \,dV\right|^2}{VB_0^2\iiint_V  {\epsilon|\mathbf{E}_{lmn}|^2} \,dV},
 \end{equation}
 where $\epsilon$ is the relative permittivity within the cavity, and $\mathbf{B}_0$ is assumed to be spatially constant and aligned with the wires. 
 
 To ensure the highest field homogeneity inside plasma haloscopes, and thus the optimal form factor, the focus has been on the study of the fundamental TM cavity mode \cite{Wuensch_PhysRevD.40.3153, bradley_microwave_2003,asztalos_large-scale_2001,stern_avoided_2019}. 
The fundamental TM mode in a WM resonator, with WM as the cavity content, exhibits a uniform axial field profile and a sinusoidal transverse pattern due to the electric wall boundary conditions and the axial polarization of the electric field \cite{Balafendiev2022}. In wedge-shaped haloscopes, the quarter-wavelength long corrugations of the walls were utilized to achieve the field uniformity in the axial direction \cite{PhysRevApplied.21.L041002, Kuo_2020}. Meanwhile, 
improving field homogeneity in WM resonators along the transversal section remains an unexplored area. 

In this work we show that placing the walls with the quarter-wavelength gap from the WM interface, we effectively transform this interface from the perfect electric conductor (PEC) into a perfect magnetic conductor (PMC). Figure~\ref{fig:sch} illustrates this concept. This transformation is similar to the conversion of a shortened transmission line into an open one by shifting the termination by an odd number of quarter-wavelengths. By modifying the boundary conditions, we create a cavity whose fundamental mode is $\mathrm{TM}_{000}$ rather than the $\mathrm{TM}_{110}$ mode of the regular cavity. This transition significantly improves the form factor and shifts the fundamental resonance frequency down to the plasma frequency. Here, we define $\mathrm{TM}_{lmn}$ modes as those with an electric field parallel to the wire axes and a magnetic field perpendicular to them.

To investigate the enhancement of form factor in wire medium (WM) resonators, we propose an analytical model taking into account the boundary conditions at the WM interfaces separated from the cavity walls by arbitrary air gaps. Using this model, we estimate the resonance frequency and form factor of square transverse section WM cavities. We compare the analytical results with numerical simulations and present experimental and full-wave simulation data for the spectra and form factors of two WM resonator prototypes fabricated using printed circuit boards (PCBs). Our findings demonstrate a significant increase in the form factor when introducing an air gap with an optimal thickness, close to a quarter of the resonant wavelength, between the WM and the cavity walls.

\section{Analytics}
\subsection{Effective boundary condition}
 \begin{figure}[h]
\centering
\includegraphics[width=1\linewidth]{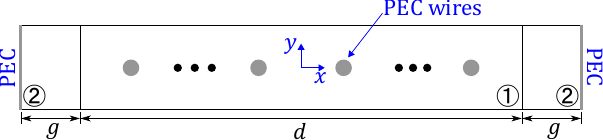}
    \caption{One-dimensional schematic of a resonator composed of a wire medium (region 1) with thickness $d$ surrounded by an air gap (region 2) with thickness $g$.}
    \label{fig:a_schemw}
\end{figure}
To assess the impact of the air gap between the wall and the WM on the modal field uniformity, we develop a one-dimensional analytical model. We consider the WM as an effective uniaxial medium, whose permittivity tensor is given by~\cite{simovski2012wire}:

\begin{equation}
\hat\epsilon=
    \begin{pmatrix}
\epsilon_t & 0 & 0 \\
0 & \epsilon_t & 0 \\
0 & 0 & \epsilon_z
\end{pmatrix}
\end{equation} 
We focus on TM modes, since TE modes weakly interact with the WM when the wire radius, $r$, is much smaller than the lattice period, $a$, i.e., $r \ll a$, for which $\epsilon_t \approx 1$. The TM modes have an electric field parallel to the wire axes ($z$-axis) and a magnetic field perpendicular to them ($y$-axis).
  We divide the system into two regions, as depicted in Fig. \ref{fig:a_schemw}. Region 1 contains the lossless (PEC wires) WM characterized by a relative permittivity~\cite{simovski2012wire}

\begin{equation}\label{eq:epsilon}
    \epsilon\equiv\epsilon_z|_{k_z=0}=1-\left(\tfrac{k_p}{k_0}\right)^2,
\end{equation} 
where $k_0$ is the vacuum wave number, $k_z$ is the $z$-component of the wave vector, and $k_p=\tfrac{2\pi}{\lambda_p}$, with $\lambda_p$ being the plasma wavelength. Region 2 represents the air gap between the WM and the PEC walls. Horizontal lines in Fig.~\ref{fig:a_schemw} outline a unit cell of the infinitely extended parallel-plate cavity. By imposing the inherent odd symmetry of the electric field and even symmetry of the magnetic field due to the PECs, we derive the TM mode solutions to Maxwell's equations as follows:
\begin{equation} \label{eq:Esols}
   E_{z}(x)= \begin{cases}
            A\cos{(k_{x1} x)} & |x|\leq \frac{d}{2} \\[0.5em]
            B\sin{\left(k_{x2}\left(\tfrac{d}{2}+g-|x|\right)\right)}  & |x|>\frac{d}{2}
        \end{cases}, 
\end{equation}

\begin{equation}\label{eq:Hsols}
H_{y}(x)= \begin{cases}
            -jA\frac{k_{x1}}{\eta_0k_0}\sin{(k_{x1} x)} & |x|\leq \frac{d}{2} \\[0.5em]
            -j\frac{x}{|x|}B\frac{k_{x2}}{\eta_0 k_0}\cos{\left(k_{x2}\left(\tfrac{d}{2}+g-|x|\right)\right)}  & |x|>\frac{d}{2}
        \end{cases},
\end{equation}
\noindent
where $k_{x1}=\sqrt{\epsilon}k_0$, and $k_{x2}=k_0$, assuming no field dependence on $y$ and $z$. We evaluate the field solutions in Eqs. \eqref{eq:Esols} and \eqref{eq:Hsols} at $x= \tfrac{d}{2}^+$ (on the side of the air gap) to derive a surface impedance that represents the effective boundary condition for the WM interface,

 \begin{equation}\label{eq:impedance}
     Z=\frac{E_z(\tfrac{d}{2}^+)}{H_y(\tfrac{d}{2}^+)}=\frac{j\eta_0 k_0}{k_{x2}}\frac{ \sin{(k_{x2}g)}}{\cos{(k_{x2}g)}}.
 \end{equation}
 This impedance vanishes for $g=0$, corresponding to a perfect electric conductor, and approaches infinity as $g$ tends to $\lambda/4$ matching the wave impedance of a perfect magnetic conductor.

\subsection{Fundamental mode of WM resonator}

In this section, we investigate the TM modes of a 1D WM resonator illustrated by Fig. \ref{fig:a_schemw} for the limit cases of $g=0$ and $g=\lambda/4$. We assume no field variation along the $z$-axis for the fundamental WM resonator mode (i.e. $k_z=0$) and consider a square resonator cross-section ( implying $k_x=k_y=k_{t1}$). Using these assumptions, we express the electromagnetic field within the effective WM as

\begin{equation} \label{eq:fields}
\begin{rcases*}
    E_z=E_0 \cos{(k_{t1}x)}\cos{(k_{t1}y)}\\
    H_x=jE_0\frac{k_{t1}}{\eta_0 k_0}\cos{(k_{t1} x)\sin{(k_{t1} y)}}\\
    H_y=-jE_0\frac{k_{t1}}{\eta_0 k_0}\sin{(k_{t1} x)\cos{(k_{t1} y)}}
\end{rcases*} |x|,\:|y| \leq \tfrac{d}{2}.
\end{equation}

For the fundamental TM mode within the effective WM region, defined by $|x|,\:|y|\leq \tfrac{d}{2}$, we define the wave number as

\begin{equation}\label{eq:kt1}
\epsilon k_0^2=k_x^2+k_y^2=2k_{t1}^2,
\end{equation}
while we define the wave number in the air gap region, for instance where $x>\tfrac{d}{2}, |y|\leq \tfrac{d}{2}$, as
\begin{equation}\label{eq:kt2}
    k_0^2=k_{t1}^2+k_{x2}^2=k_{t1}^2+k_{t2}^2.
\end{equation}
Utilizing Eqs. \eqref{eq:kt1} and \eqref{eq:kt2}, and solving for $k_{t1}$ and $k_{t2}$,
\begin{equation} \label{eq:kts}
\begin{split}
    k_{t1}^2=\frac{\epsilon}{2}k_0^2, \\
    k_{t2}^2=\left(1-\frac{\epsilon}{2}\right)k_0^2.
\end{split}
\end{equation}

It is important to note that in this model, the WM resonator has no corners, which may have impact for the real resonator. Despite this approximation, we can employ the one-dimensional solution to estimate the two-dimensional behavior, analogous to Marcatili's method for dielectric rectangular waveguides \cite{Marcatili1969}. This approximation is valid when the field is primarily concentrated in the central region.

Matching the surface wave impedance of the field at the effective WM boundary (Eqs. \eqref{eq:fields}) with that of the effective boundary condition (Eq. \eqref{eq:impedance}), particularly at $x=\tfrac{d}{2}$, $|y|\le \tfrac{d}{2}$, we derive the following dispersion equation:

\begin{equation}\label{eq:dispersion}
    k_{t1}\tan{\left(k_{t1}\tfrac{d}{2}\right)}-k_{t2}\cot{(k_{t2}g)}=0.
\end{equation}
This dispersion equation yields the resonance frequencies of the TM modes. Notably, using Eq.~\eqref{eq:kts}, we can determine the resonance frequency of the fundamental TM mode. For the fundamental mode, when $g=\lambda/4$, the second term in Eq.~\eqref{eq:dispersion} vanishes, implying that $k_{t1}=0$. Consequently, $\epsilon=0$ and $k_0=k_p$, leading to a constant electric field according to Eq.~\eqref{eq:fields}. It is worth mentioning that Eq.~\eqref{eq:dispersion} is similar to the dispersion equation derived in Ref.~\cite{mett_axially_2001}. However, while Ref.~\cite{mett_axially_2001} achieves field homogeneity in a single direction by using a central metallic waveguide operating at its cutoff frequency, our approach requires a WM in the central region of the cavity operating at the plasma frequency to ensure field homogeneity in both transverse directions. 

In order to quantify the field uniformity in the WM resonator, we use a 2D version of the form factor in Eq.~\eqref{Ffactor},
\begin{equation}
    C = \frac{\left(\iint_S  {E_z} \,dS\right)^2}{S\iint_S  {|\mathbf{E}|^2} \,dS},
\end{equation}
where $S$ is the area of the transversal section of the 2D WM resonator, $(d+2g)\times (d+2g)$. 

\subsection{Results}

\begin{figure}[h]
\centering
\includegraphics[width=1\linewidth]{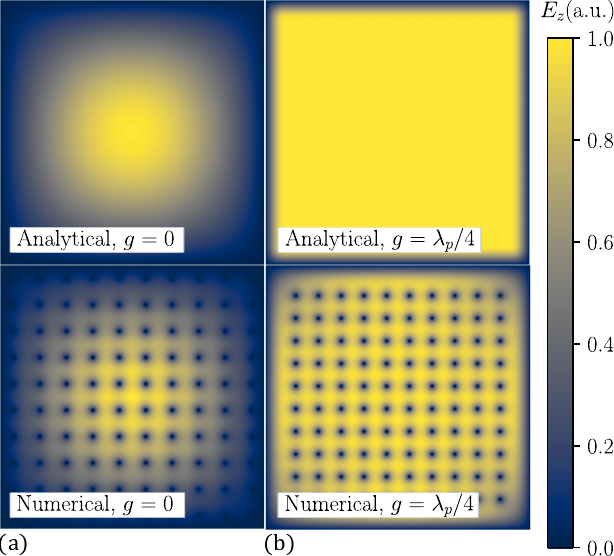}
    \caption{Electric field distribution in a wire medium (WM) resonator: (a) Regular WM resonator with no air gap ($g=0$), and (b) WM resonator with an air gap of thickness $g = \lambda_p/4$. The introduction of the air gap ($g = \lambda_p/4$) results in a more uniform electric field distribution.}
    \label{fig:a_field2D}
\end{figure}

\begin{figure}[h]
\centering
\includegraphics[width=1\linewidth]{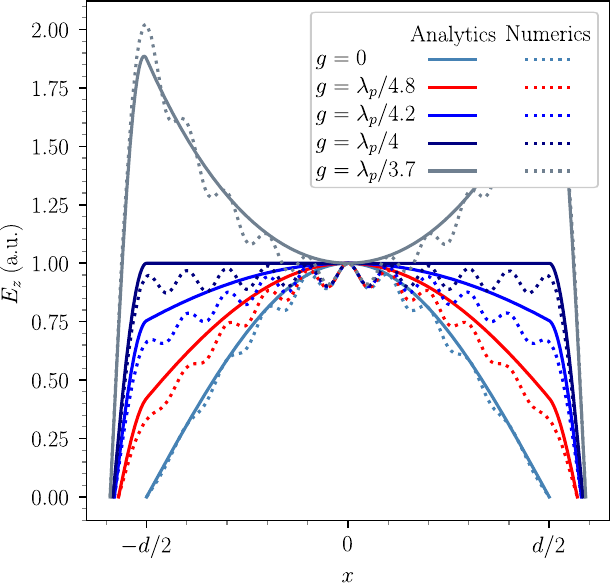}
    \caption{Wire medium (WM) resonator electric field profile for different air gap thicknesses, $g$. The WM extends from $x=-\tfrac{d}{2}$ to $x=\tfrac{d}{2}$. A uniform electric field in the WM region is observed at $g = \lambda_p/4$.}
    \label{fig:a_field}
\end{figure}

\begin{figure}[h]
\centering
\includegraphics[width=1\linewidth]{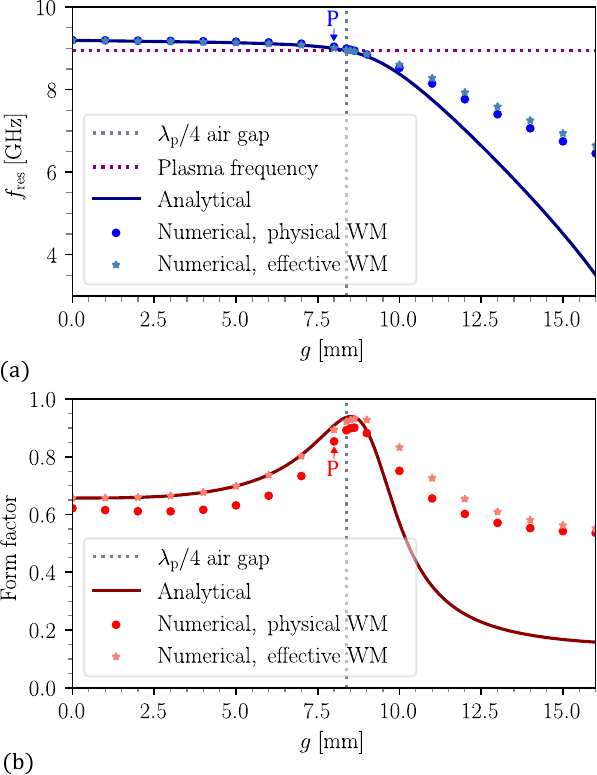}
    \caption{Characteristics of wire medium (WM) resonator fundamental TM mode. a) Resonance frequency and b) form factor for different air gap thicknesses, $g$. A form factor of approximately 0.92 is observed at $g = \lambda_p/4$ for an effective WM (theory and numerics) while a form factor of 0.89 is observed for the physical WM (numerics). Point P corresponds to the air gap thickness used in the experimental section.}
    \label{fig:ff_fr}
\end{figure}
We used full-wave simulations in a COMSOL Multiphysics eigenmode solver to validate the applicability of our 1D theory to a 2D cavity. In both our analytical calculations and simulations, the cavity comprised an array of 10$\times$10 infinitely long wires with wire radius $r$~=~0.5~mm and inter-wire spacing $a$~=~10~mm. We adopted the generalized quasi-static approach from Ref.~\cite{Kumar201243} to determine the requisite plasma frequency for the theoretical model, which expresses the plasma frequency as a function of $r$ and $a$.

Figure~\ref{fig:a_field2D} illustrates the arbitrary normalized electric field distribution for the fundamental TM mode over the transverse ($xy$) plane of the cavity for two specific cases: $g=0$ and $g=\lambda_p/4$. Our numerical and analytical findings confirm that introducing a $\lambda_p/4$ air gap homogenizes the electric field distribution in a finite-cross section sample of the WM.

The electric field profile along the $x$-axis for the fundamental TM mode of a WM resonator varies with the gap width, $g$, as shown in Fig.~\ref{fig:a_field}. The electric field homogeneity improves with increasing $g$ until reaching a maximum at $g=\lambda_p/4$. Introducing a quarter-wavelength air gap between the effective wire medium and the PEC walls significantly flattens the electric field profile, confining the field decay primarily to the newly added $\lambda_p/4$ region. This coincides with a pronounced magnetic field enhancement within the same gap, offering potential for readout applications. For the gap exceeding $\lambda_p/4$, the field concentrates near the resonator walls, and the underlying assumption of field confinement within the wire medium becomes not adequate.

Figure~\ref{fig:ff_fr} illustrates the resonant frequency (Fig.~\hyperref[fig:ff_fr]{5a}) and form factor (Fig.~\hyperref[fig:ff_fr]{5b}) of the fundamental mode in such WM resonator as a function of the air gap, $g$. The resonant frequency remains nearly constant for $0\leq g\leq\lambda_p/4$, aligning with the plasma frequency at $g=\lambda_p/4$. Conversely, the form factor progressively increases, reaching a maximum at an air gap of approximately $g=\lambda_p/4$. While the resonant frequency exhibits excellent agreement between analytical and numerical results, coming from both a physical WM and an effective WM, within this range, a slight discrepancy arises in the form factor. This error results from the evident difference between the effective and physical (real) WM. In the effective-medium model there are no wires and the field does not vanish inside this medium, whereas in the physical WM it vanishes within every wire -- see the dark spots in the bottom panel of Fig.~\hyperref[fig:a_field2D]{3b}. 

In accordance with numerical calculations of the mode field of our WM resonator the maximally achievable form factor corresponding at 8.9 GHz to $g=\lambda_p/4=$8.4 mm is equal to 0.92, as well as in the analytical model. However, the simulation of the physical WM corresponding to the bottom panel of Fig.~\hyperref[fig:a_field2D]{3b} shows the form factor 0.89. Despite this, the maximum form factor occurs at nearly the same air gap value in both numerical simulations and the analytical model. Our strategy considerably enhances the form factor of microwave cavities for axion detection, surpassing the maximum form factor of 0.69 found in cylindrical cavities \cite{asztalos_large-scale_2001, bradley_microwave_2003}. 

For smaller gaps the agreement between the analytical and numerical results corresponding to physical WM is good, and is excellent for the numerical simulations of the effective WM sample surrounded by the PEC walls. Thus, these results fully justify the model of an effective WM sample for $g\le\lambda_p/4$.
For $g>\lambda_p/4$, the agreement between the analytical model and numerical simulations disappears due to the effect of the field concentration in the air gap. This effect invalidates the initial assumption of field confinement within the WM sample. However, the agreement between the numerical results obtained for the effective WM sample and for the array of separate wires keeps.

It is evident that the maximal form factor corresponds to the perfect magnetic wall effectively located on the WM-air interface. The PEC wall shifts the phase of a reflected wave by $\pi$, the air gap grants an extra $\pi$, so the phase of the reflected wave does not change. Therefore, the mode inside the WM sample is the standing wave shaped by four perfect magnetic boundaries surrounding the sample. Note that this effective boundary condition can be achieved through various methods that introduce an extra phase shift of $\pi$ between incident and reflected waves, as detailed in \cite{mett_axially_2001,anderson_cavities_2002,hyde_cavities_2002}. This flexibility may help the TM$_{000}$ modes implementation in innovative plasma haloscope designs that require tunability.

\begin{figure}[h]
\centering
{\includegraphics[width=1\linewidth]{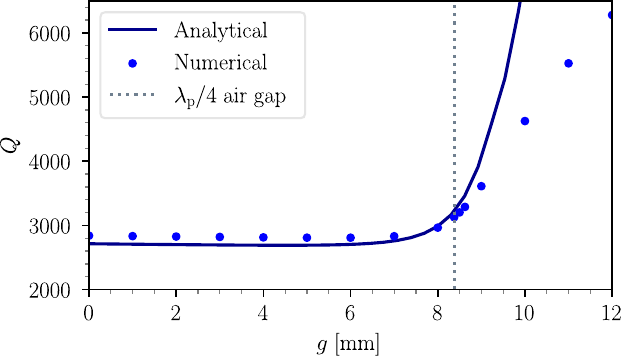}}
    \caption{Quality factor of wire medium (WM) resonator fundamental TM mode for different air gap thicknesses, $g$. Wires were modeled as copper with a conductivity of $\sigma=5.7\times 10^7~\mathrm{S/m}$.}
    \label{fig:qf}
\end{figure}

To analyze the quality factor, we focused solely on losses due to the wires, as they are the dominant loss mechanism in WM cavities~\cite{Balafendiev2022}. We considered copper wires with a conductivity of $\sigma = 5.7 \times 10^7~\mathrm{S/m}$ for both the analytical model and the 2D numerical simulations. By using the formulas for the quality factor described in Ref.~\cite{Balafendiev2022}, along with the updated plasma frequency and wire inductance from Ref.~\cite{Kumar201243}, we obtained the analytical results. Introducing an air gap of a quarter-wavelength in the wire media resonator significantly enhances the form factor and slightly increases the quality factor, as shown in Fig.~\ref{fig:qf}. When the air gap exceeds a quarter-wavelength, the quality factor rapidly increases as the field becomes primarily concentrated in the newly added air gap region.
 
%

\section{Experiment}
\begin{figure}[h]
\centering
\includegraphics[width=1\linewidth]{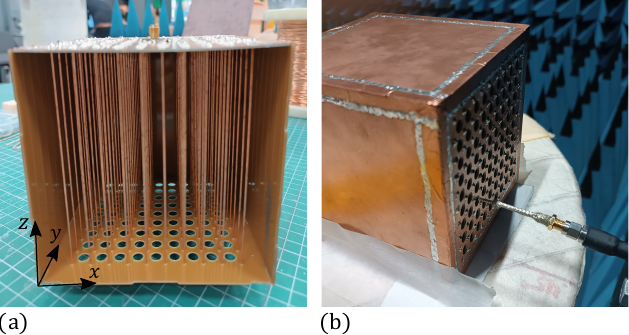}
    \caption{ Prototype WM resonators with different wall spacing. The resonators consist of 10 x 10 arrays of copper wires with a period of $a=1$ cm. The first resonator is a regular resonator with a distance of $a/2$ between the wires at the edge and the walls, and the second one is $\lambda/4.2$-shifted with a distance of approximately $a/2+\lambda/4.2$ between the wires at the edge and the walls. (a) Unfinished state of the regular resonator. Copper wires (radius 0.5 mm) are inserted into the FR-4 PCB walls with a period of 1~cm.
(b) Finished prototype being scanned through pre-drilled holes by a movable antenna.}
    \label{fig:e_prototypes}
\end{figure}
We fabricated two prototypes to validate the claimed effect. The first prototype represents the regular (conventional) configuration, see for example Ref.~\cite{Balafendiev2022}, where the air gap between the array of wires and the cavity walls is equal $a/2$. The second prototype incorporates a configuration we specially optimized for the cubic resonator to enhance field homogeneity and to avoid situations where the air gap exceeds $\lambda_p/4$, as this would lead the field to localize mainly near the cavity walls (see Fig.~\ref{fig:a_field}). In this 3D case, the designed gap is slightly smaller than the quarter-wavelength. So, for the optimized resonator, the edge wires of the array are distanced from the walls by ($a/2 + \lambda_p/4.2$). For the 2D resonator, this configuration would correspond to point P in Fig.~\ref{fig:ff_fr}. Below we call our optimized cubic resonator the $\lambda_p/4.2$-shifted one.

We fabricated the prototype enclosures using single-sided metal FR-4 PCBs of $1$~mm thickness. We arranged these PCBs into cubes with dimensions of 10~cm x 10~cm x 10~cm for the regular resonator and 11.6~cm x 11.6~cm x 11.6~cm for the $\lambda/4.2$-shifted resonator. To complete the resonators, we inserted and welded one hundred copper wires, each with a radius of 0.5 mm, into the PCB cubes with a periodic spacing of $a = 1$ cm, as shown in Fig.~\hyperref[fig:e_prototypes]{6a}.

We equipped each resonator prototype with a 4-mm long SMA connector attached to its bottom plate for signal feeding. The SMA port functions similarly to a monopole antenna, exciting the TM modes within the resonator cavity. To characterize the electric field distribution of the TM modes, we positioned a separate monopole antenna through a pre-drilled array of holes on the top plate of each resonator, as illustrated Fig.~\hyperref[fig:e_prototypes]{6b}. This monopole antenna was a coaxial cable with a center conductor radius of 0.25 mm, extending 4 mm into the resonators through the holes. The hole array comprised 81 individual holes, arranged in a square grid with a period of $a = 1$ cm and a hole radius of $r_h = 3$ mm.
\begin{figure}[h]
\centering
{\includegraphics[width=1\linewidth]{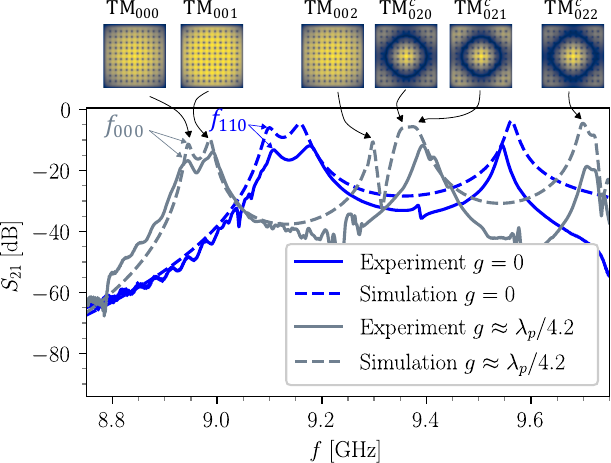}}
    \caption{Experimentally measured $\mathrm{S}_{21}$ parameter for wire media resonators: (a) Regular resonator and (b) $\lambda_p/4.2$-shifted resonator. Measurements were performed by placing the movable monopole antenna in the central hole. The resonance frequency of the fundamental TM mode in the regular resonator ($f_{110}$) is $9.11$ GHz (experiment) and $9.10$ GHz (simulation), while in the $\lambda_p/4.2$-shifted resonator ($f_{000}$) it is $8.943$ GHz (experiment and simulation). The insets show $|E_z|$ in the $xy$ plane (as in Fig.~\ref{fig:sch}) at $10~\mathrm{mm}$ above the center of the $\lambda_p/4.2$-shifted resonator. In the color maps, yellow represents 1 and dark blue represents 0 in arbitrary units (a.u.) for $|E_z|$.} 
    \label{fig:e_spara}
\end{figure}

We measured the S-parameters using a Rohde $\&$ Schwarz ZVB20 vector network analyzer (VNA) across a frequency range of 8.75-9.75 GHz. The fixed antenna, located on the bottom plate of each prototype, was connected to one port of the VNA, while the movable antenna on the top plate was connected to the other port. To measure the $\mathrm{S_{21}}$, we positioned the monopole antenna in the array of holes. Fig.~\ref{fig:e_spara} displays the measured $\mathrm{S_{21}}$  when the movable antenna was placed in the central hole of each prototype. We observe that the spectrum of the $\lambda_p/4.2$-shifted resonator is red-shifted relative to that of the regular resonator, as predicted by theory for the fundamental mode. In general, the $\mathrm{TM}_{lmn}$ modes in the spectrum of the $\lambda/4.2$-shifted resonator are red-shifted compared to those of the regular resonator, allowing the $\mathrm{TM}_{02n}^c$ modes to fall within the studied frequency range, as detailed in Appendix A. The superscript $c$ in the $\mathrm{TM}_{02n}^c$ modes indicates a cross-coupling between the $x$ and $y$ dependence of the fields, as discussed in Appendix A. 

We conducted full-wave simulations of cubic resonators, corresponding to the experimental prototypes, using CST Studio. Copper was modeled with a conductivity of $\sigma=5.7\times 10^7\; \mathrm{S/m}$, while the FR-4 layers of the PCBs were represented as lossy dielectrics with a dielectric constant of $\varepsilon=4.3$ and a loss tangent of $\tan{\delta}=0.025$. We measured the $\mathrm{S_{21}}$ parameter and positioned a field monitor at the first resonance frequency for each resonator.

\begin{figure}[h]
\centering
\includegraphics[width=1\linewidth]{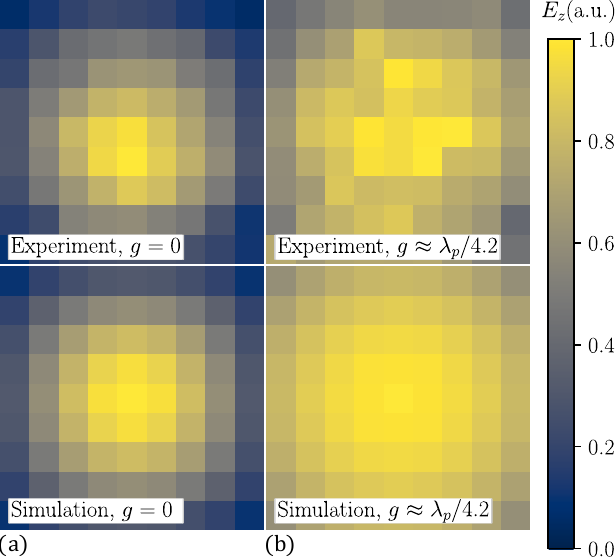}
    \caption{Measured electric field distribution of the fundamental TM mode in wire media resonators: (a) Regular resonator and (b) $\lambda/4.2$-shifted resonator. Pixel intensities represent the magnitude of the $S_{21}$ parameter measured at the fundamental resonance frequencies, $f_{110}$ and $f_{000}$ (marked in Fig.~\ref{fig:e_spara}), through pre-drilled holes at one end of the cavities.}
    \label{fig:e_field}
\end{figure}

We characterized the operation of the two resonators under comparison by examining the first peaks of the $\mathrm{S_{21}}$-parameter spectra. These peaks correspond to the $\mathrm{TM_{110}}$ mode for the regular resonator and the quasi-$\mathrm{TM_{000}}$ mode for the $\lambda_p/4.2$-shifted resonator, as shown in Fig. \ref{fig:e_spara}. We do not consider high-order modes, as they are not relevant for our purpose of achieving the highest field homogeneity (see e.g. in Refs.~\cite{anderson_cavities_2002, Balafendiev2022}). Fig.~\ref{fig:e_field} presents the measured electric field distributions for both prototypes at the corresponding frequencies ($f_{000}$ for the $\lambda_p/4.2$-shifted resonator and $f_{110}$ for the regular one). We observe a significant improvement in field homogeneity for the $\lambda_p/4.2$-shifted resonator compared to the regular resonator. This observation is further supported by the measured form factors, with a value of $0.77$ (experiment), $0.80$ (simulation) for the $\lambda_p/4.2$-shifted resonator compared to $0.61$ (experiment), $0.64$ (simulation) for the regular resonator. The regular resonator consists of a cavity filled exclusively with WM, as described in Ref. \cite{Balafendiev2022}.

So, our experimental results demonstrate an improvement in field homogeneity due to the optimized gap between the WM sample and the resonator walls. Slight discrepancies between experimental and simulation results seen in the field maps of Fig.~\ref{fig:e_field} can be attributed to experimental limitations in the prototype fabrication and the field scanning procedure. Due to manual soldering of the wires, variations in wire length within the prototypes may exist. Additionally, excessive heat during soldering could have caused material deformations, leading to bending in either the wires or PCB plates. Moreover, the scanning procedure, involving the antenna traversing through holes, might have introduced slight variations in field measurements if the antenna was not perfectly centered at each hole. This could explain the difficulty in differentiating between high field levels and the exact location of the maximum field, which was not precisely at the central hole as simulations predicted.

\section{Conclusions}
We introduced a new method to enhance field homogeneity within WM resonators. Our theoretical model, supported by numerical simulations, demonstrated that properly increasing the air gap between the resonator walls and the WM sample, one can effectively manipulate the field homogeneity. When the air gap reached one-quarter of the resonant wavelength, the field homogeneity peaked, theoretically resulting in a form factor of 0.92 for an effective WM model (theory and simulations) and 0.89 for a physical array of wires (simulations). Additionally, this enhancement in the form factor was accompanied by a shift in the resonant mode frequency toward the plasma frequency, along with a slight increase in the quality factor.

We experimentally validated the proposed method by developing two prototype resonators: a regular WM resonator containing the WM sample with the conventional half-period air gap between the wires and the walls and an optimized $\lambda_p/4.2$-shifted resonator. We devised a method to measure the field distribution within the transverse section of the cavities, enabling the subsequent calculation of the form factor. This method involved measuring the magnitude of the $S_{21}$ parameter using a fixed antenna at one end of the resonator and a scanning antenna at the other end. The scanning antenna traversed predrilled holes within the resonator. Comparing experimental results with full-wave simulations, we found that the form factor in a standard WM resonator was 0.61 (experiment) and 0.64 (simulation), whereas in the approximately $\lambda/4.2$-shifted WM resonator, the form factor increased to 0.77 (experiment) and 0.80 (simulation). Our findings validate and substantiate using the TM$_{000}$ mode to improve the form factor in plasma haloscopes, opening new avenues for WM cavity applications where precise control of the internal field profile is essential.

In the case of a dark matter search, the signal power is directly proportional to the form factor. If one achieves the maximum improvement of~$40\%$ in the form factor the scanning rate of the experiment, which goes as the square of the power, is improved by almost a factor of two, especially when one considers the increased quality factor as well. This improvement comes at no real additional costs, unlike say increasing the size or strength of the magnetic field. Further, these improvements are not mutually exclusive with other enhancements such as higher magnetic fields or lower noise floors. Particularly at high frequencies improving the scanning speed of dark matter experiments is crucial to search for dark matter.

In order to realize a practical experiment, there are two main avenues for future work. The first is to adapt the cavity design to have a cross-section better suited for a cylindrical bore of a solenoidal magnet. Secondly, given purpose of the plasma haloscopes inherently requires their resonance frequency to be tunable in order to search for dark matter at different frequencies, we plan to investigate the effectiveness of the proposed approach in improving the form factor when the mode frequency varies. This will involve considering both operation away from the quarter-wave condition and adjustment of the walls to maintain it, similarly to the tuneable rectangular cavity in Ref.~\cite{McAllister2024}.

\section*{Acknowledgements}
The authors thank Eugene Koreshin for helpful discussions and also thank the members of the ALPHA Consortium for discussion and support.

Fermilab is operated by the Fermi Research Alliance, LLC under Contract DE-AC02-07CH11359 with the U.S. Department of Energy.

\section*{Data availability}
The data that support the findings of this article are
openly available \cite{enriquez_2025_15295850}.

\appendix
\section{Higher order modes}
\begin{figure}[h]
\centering
\includegraphics[width=1\linewidth]{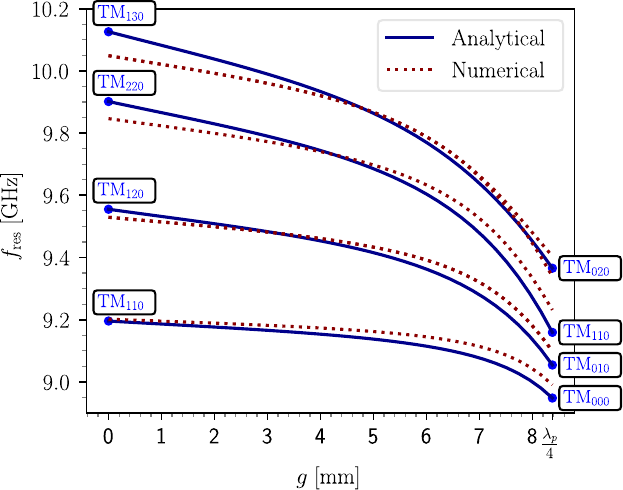}
    \caption{Resonant frequencies of TM modes in WM cavities with varying air gaps ($g$). For $g = 0$, the TM modes are bounded by perfect electric conductors (PEC), while at $g = \lambda_p/4$, they experience perfect magnetic conductor (PMC) boundary conditions. The branches of the $\mathrm{TM}_{010}$ and $\mathrm{TM}_{020}$ modes also include the degenerate modes, $\mathrm{TM}_{100}$ and $\mathrm{TM}_{200}$, respectively.}
    \label{fig:modes_vs_g}
\end{figure}
For the purposes of dark matter searches, the TM$_{000}$ mode is most uniform and so gives the the highest form factor $C$. However, higher order modes also exist, and may be useful for gravitational wave searches~\cite{Berlin:2021txa,capdevilla2024gravitational}. Electromagnetic waves with an electric field polarized along the wire axis ($z$-axis) experience an effective relative permittivity for the WM, given by~\cite{simovski2012wire}

\begin{equation}\label{eq:ap_epsilon}
    \epsilon_{z}=1-\frac{k_p^2}{k_0^2-k_z^2},
\end{equation}
when field variations along $z$ are allowed, i.e., when $k_z\geq0$. 

\begin{figure*}[t]
\centering
\includegraphics[width=1\linewidth]{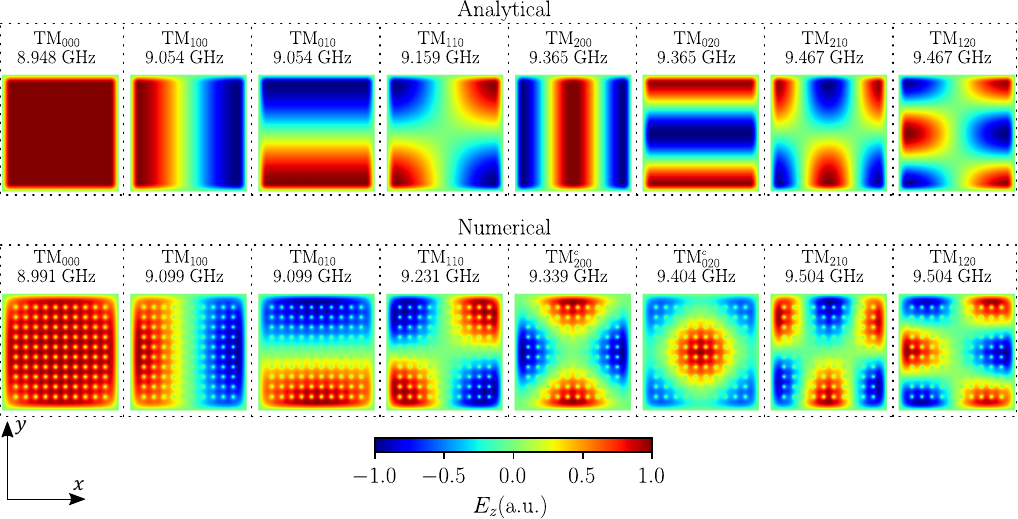}
    \caption{Electric field distributions for the \( \mathrm{TM}_{lm0} \) modes in a $\lambda/4$-shifted wire-medium (WM) resonator, where the modes experience an effective perfect magnetic conductor (PMC) boundary at the WM-air gap interface.
}
    \label{fig:modes_field}
\end{figure*}

In general, we express the cavity mode solutions as either even or odd. For the WM region, these solutions take the form:
\begin{equation} \label{eq:ap_fields}
\begin{rcases*}
    E_z^{lmn}=E_0 f_x{(k_{x1}^lx)}g_y{(k_{y1}^my)}h_z{(k_z^nz)}\\
    H_x^{lmn}=-jE_0\frac{1}{\eta_0 k_0}f_x{(k_{x1}^lx)}g^{\prime}_y{(k_{y1}^my)}h_z{(k_z^nz)}\\
    H_y^{lmn}=jE_0\frac{1}{\eta_0 k_0}f^{\prime}_x{(k_{x1}^lx)}g_y{(k_{y1}^my)}h_z{(k_z^nz)}
\end{rcases*} |x|,\:|y| \leq \tfrac{d}{2},
\end{equation}
where the functions $f_x, g_y, h_z$ are either $\sin$ or $\cos$ functions depending on the integer numbers $l, m, n$, respectively. When $l, m,$ or $n$ is zero or an even number, the associated function is a $\cos$ function; otherwise, it is a $\sin$ function.

The $z$-component of the wavevector, $k_z^n$, is not affected by the newly added effective boundary condition, so it remains $k_z^n=\tfrac{n\pi}{d}$~\cite{Balafendiev2022}. The components $k_x^l$ and $k_y^m$ are discretized to satisfy the effective boundary condition determined by the impedance in Eq.~\eqref{eq:impedance}, which leads to the dispersion equation Eq.~\eqref{eq:dispersion} for even functions. On the other hand, for odd functions, it results in  

\begin{equation}\label{eq:ap_dispersion_odd}
   k_{t1}\cot{\left(k_{t1}\tfrac{d}{2}\right)}+k_{t2}\cot{\left(k_{t2}g\right)}=0,
\end{equation}  
where the subscript $t$ corresponds to either $x$ or $y$.  

In addition, the wave number in each region satisfies the following dispersion relations: 

In the wire medium, for $|x|,|y|\leq \tfrac{d}{2}$,  
\begin{equation}\label{eq:ap_k_wm}
    \epsilon_z(k_0^{lmn})^2=(k_{x1}^l)^2+(k_{y1}^m)^2+\epsilon_z\left(\frac{n\pi}{d}\right)^2.
\end{equation}  

In the air gap, for $|x|>\tfrac{d}{2}$ and $|y|\leq\tfrac{d}{2}$,  
\begin{equation}\label{eq:ap_k_gap1}
    (k_0^{lmn})^2=(k_{y1}^m)^2+(k_{x2}^l)^2+\left(\frac{n\pi}{d}\right)^2.
\end{equation} 

In the air gap, for $|y|>\tfrac{d}{2}$ and $|x|\leq\tfrac{d}{2}$,  
\begin{equation}\label{eq:ap_k_gap2}
    (k_0^{lmn})^2=(k_{x1}^l)^2+(k_{y2}^m)^2+\left(\frac{n\pi}{d}\right)^2.
\end{equation}  

Solving for $k_{x1}^m$ in Eq.~\eqref{eq:ap_k_wm} and substituting the result into Eq.~\eqref{eq:ap_k_gap1}, along with the relative permittivity from Eq.~\eqref{eq:ap_epsilon}, we obtain:  
\begin{equation}\label{eq:ap_kt2_kt1}
    k_{t2}^2 = k_p^2 + k_{t1}^2,
\end{equation}  
where the subscript $t$ is either $x$ or $y$.  

Thus, by using the dispersion equations (Eqs.~\eqref{eq:dispersion} or \eqref{eq:ap_dispersion_odd}) along with Eq.~\eqref{eq:ap_kt2_kt1}, we determine the $k_{x1}^l$ and $k_{y1}^m$ components for the specified $l$ and $m$ orders. Once we establish the order $n$ of the solution, we fully solve for the eigenvalue $k_0^{lmn}$ using the dispersion relation in Eq.~\eqref{eq:ap_k_wm}.



Figure \ref{fig:modes_vs_g} shows the resonance frequency of the $\mathrm{TM}_{lm0}$ modes as a function of the air gap ($g$). Analytical and numerical results indicate that the spectrum of modes of the $\lambda/4$-shifted resonator ($g=\lambda_p/4$) is compressed compared to that of the regular resonator ($g=0$). In particular, it explains the apparition of the $\mathrm{TM}_{020}^c$ modes for the spectrum of the $\lambda/4.2$-shifted resonator, as shown in Fig. \ref{fig:e_spara}. The $\mathrm{TM}_{020}^c$ has a cross coupling in the $x$- and $y$-dependence of the fields which explains the breaking of the degeneracy between the $\mathrm{TM}_{020}$ and the $\mathrm{TM}_{200}$ modes. This effect is evident in the numerical results, which show two distinct branches around the analytical $\mathrm{TM}_{020}$ mode.

Figure \ref{fig:modes_field} shows the first $ \mathrm{TM}_{lm0} $ modes in a $ \lambda/4 $-shifted cavity. We can see that the effective PMC boundary condition, introduced by including an air gap of length $ g = \lambda_p/4 $ in WM cavities, applies not only to the fundamental mode but also to higher-order modes, with almost all the modes being predicted by the analytical calculation. 

The exceptions to this agreement are the the $\mathrm{TM}_{200} $ and $ \mathrm{TM}_{020} $ modes. The mismatch is not due to the effective PMC description breaking down, but because $\mathrm{TM}_{200} $ and $ \mathrm{TM}_{020} $ modes are degenerate in the analytical model, which does not include cross-coupling effects. These effects arise from the neglected corners of the cavity. According to numerical results, these modes should actually be denoted as $ \mathrm{TM}_{200}^c $ and $ \mathrm{TM}_{020}^c $, where the superscript $c$ indicates cross-coupling between the $ x $- and $ y $-dependence of the field. This cross-coupling leads to an eigenvalue splitting of the expected degenerate modes $ \mathrm{TM}_{200}$ and $ \mathrm{TM}_{020} $. Specifically, the $\mathrm{TM}_{200}^c$ mode results from the superposition of the $\mathrm{TM}_{200}$ and the $\mathrm{TM}_{020}$ modes, while the $\mathrm{TM}_{020}^c$ consists of the $\mathrm{TM}_{200}$ mode and a $\pi$-shifted $\mathrm{TM}_{020}$ mode. One way to properly estimate the frequency splitting of the cross-coupled modes is to apply a perturbative approach, incorporating coupled mode theory into the analytical model. This method has shown improvements over Marcatilli's approach for studying rectangular waveguides \cite{pollock2003integrated}.

\section*{References}

\end{document}